\documentclass[pdflatex,sn-nature]{sn-jnl}
\usepackage{amsmath}
\usepackage{siunitx}
\usepackage{cleveref}
\usepackage{CJKutf8}
\usepackage{rotating}
\usepackage{booktabs}
\usepackage{subfig}
\usepackage{ulem}
\usepackage{textcomp}
\usepackage{xcolor,textcomp}
\usepackage{graphicx}
\usepackage{txfonts}
\usepackage{placeins}

\usepackage{etoolbox}

\makeatletter
\patchcmd{\@printaffiliations}{\space.}{.}{}{}
\makeatother
\begin{document} 

\title{A Detection of Sulfur-Bearing Cyclic Hydrocarbons in Space}


\date{to be submitted}
\author*[1]{\fnm{Mitsunori} \sur{Araki}}\email{araki@mpe.mpg.de}
\author[2]{\fnm{Miguel} \sur{Sanz-Novo}}
\author[1]{\fnm{Christian P.} \sur{Endres}}
\author[1]{\fnm{Paola} \sur{Caselli}}
\author[2]{\fnm{V\'{i}ctor M.} \sur{Rivilla}}
\author[2]{\fnm{Izaskun} \sur{Jim\'{e}nez-Serra}}
\author[2]{\fnm{Laura} \sur{Colzi}}
\author[3]{\fnm{Shaoshan} \sur{Zeng}}
\author[2]{\fnm{Andr\'{e}s} \sur{Meg\'{i}as}}
\author[2]{\fnm{\'{A}lvaro} \sur{L\'{o}pez-Gallifa}}
\author[2]{\fnm{Antonio} \sur{Mart\'{i}nez-Henares}}
\author[2,4]{\fnm{David} \sur{San Andr\'{e}s}}
\author[5,6]{\fnm{Sergio} \sur{Mart\'{i}n}}
\author[7]{\fnm{Miguel A.} \sur{Requena-Torres}}
\author[8]{\fnm{Juan} \sur{Garc\'{i}a de la Concepci\'{o}n}}
\author*[1]{\fnm{Valerio} \sur{Lattanzi}}\email{lattanzi@mpe.mpg.de}
\affil[1]{
  \orgdiv{Center for Astrochemical Studies}
  \orgname{Max-Planck-Institut f\"ur extraterrestrische Physik}
  \orgaddress{
    \street{Giessenbachstrasse 1}
    \city{Garching bei Munchen}
    \postcode{85748}
    \country{Germany}%
  }
}
\affil[2]{
  \orgdiv{Centro de Astrobiolog\'ia (CAB), CSIC-INTA}
  \orgaddress{
    \street{Ctra de Torrej\'on a Ajalvir, km 4}
    \city{Torrej\'on de Ardoz}
    \postcode{28850}
    \state{Madrid}
    \country{Spain}%
  }
}
\affil[3]{
  \orgdiv{Star and Planet Formation Laboratory, Cluster for Pioneering Research, RIKEN}
  \orgaddress{
    \street{2-1 Hirosawa}
    \city{Wako}
    \postcode{351-0198}
    \state{Saitama}
    \country{Japan}%
  }
}
\affil[4]{
  \orgdiv{Departamento de F\'isica de la Tierra y Astrof\'isica, Facultad de Ciencias F\'isicas, Universidad Complutense de Madrid}
  \orgaddress{
    \postcode{28040}
    \state{Madrid}
    \country{Spain}%
  }
}
\affil[5]{
  \orgdiv{European Southern Observatory}
  \orgaddress{
    \street{Alonso de C\'ordova 3107}
    \city{Vitacura}
    \postcode{763 0355}
    \state{Santiago}
    \country{Chile}%
  }
}
\affil[6]{
  \orgdiv{Joint ALMA Observatory}
  \orgaddress{
    \street{Alonso de C\'ordova 3107}
    \city{Vitacura}
    \postcode{763 0355}
    \state{Santiago}
    \country{Chile}%
  }
}
\affil[7]{
  \orgdiv{Department of Physics, Astronomy and Geosciences, Towson University}
  \orgaddress{
    \city{Towson}
    \postcode{21252}
    \state{MD}
    \country{USA}%
  }
}
\affil[8]{
  \orgdiv{Departamento de Qu\'imica Org\'anica e Inorg\'anica, Facultad de Ciencias, and IACYS-Green Chemistry and Sustainable Development Unit, Universidad de Extremadura}
  \orgaddress{
    \city{Badajoz}
    \postcode{06006}
    \country{Spain}%
  }
}
 
\abstract{%
Molecules harbouring sulfur are thought to have played a key role in the biological processes of life on Earth, and thus, they are of much interest when found in space. Here we report on the astronomical detection of a six-membered sulfur-bearing cyclic hydrocarbon in the interstellar medium. Observations of the Galactic Centre molecular cloud G+0.693-0.027 reveal the presence of 2,5-cyclohexadien-1-thione, which is a structural isomer of thiophenol ($c$-C$_6$H$_6$S). For the astronomical identification, we first performed precise laboratory measurements of the thiophenol discharge products system. These measurements, conducted in the radio band using a chirped-pulse Fourier transform microwave spectrometer, enabled us to characterize this highly polar molecular species and provided unambiguous fingerprints needed to identify this organosulfur compound in space, which now ranks as the largest interstellar sulfur-bearing molecule. These results herald the discovery of a family of prebiotically relevant sulfur-bearing species, which potentially act as a bridge between the chemical inventory of the interstellar medium and the composition of the minor bodies of the Solar System.}
\maketitle

\section{Introduction}

Understanding the origin of life remains one of the most profound and enduring challenges in science. Over many years, numerous theories have been advanced to address this question (for example, \cite{Oparin1938}). Among these, the hypothesis that life’s building blocks were synthesized in interstellar space and later delivered to the primitive Earth via comets and meteorites has garnered substantial attention \cite{DELSEMME2000,Osinski2020}. This model is supported by the detection of a rich inventory of prebiotic organic molecules in comets (for example, \cite{Crovisier2004, altwegg2016prebiotic}), meteorites (for example, \cite{Yabuta2007}), and asteroids (for example, \cite{oba2023ryugu,connolly2025overview}). 

The role of sulfur-bearing molecules is pivotal in the general discussion of prebiotic chemistry. Sulfur (S hereafter) is indispensable to many biological processes and is considered an essential element for life on Earth (for example, \cite{Todd2022}). Yet, there exists a notable discrepancy between the S compounds observed in interstellar space and those found in meteoritic material. In the gas phase of the interstellar medium (ISM), only a limited array of S species—predominantly small molecules with up to nine atoms (CH$_3$SCH$_3$  \cite{sanz2025abiotic} and C$_2$H$_5$SH \cite{kolesnikova2014spectroscopic,rodriguez-almeida2021a})—have been detected. Furthermore, whereas the observed sulfur abundance in the diffuse ISM is consistent with the cosmic value, in dense molecular clouds the overall abundance of gas-phase S species in dense molecular clouds is significantly lower (for example, one order of magnitude lower \cite{fuente2023gas}) than expected when compared with the cosmic S/H abundance ratios. This S depletion in the ISM suggests that a substantial fraction of sulfur may be sequestered in forms that are not readily detectable, such as refractory compounds or as constituents of dust grains, leaving the dominant carriers of sulfur in interstellar space unidentified.

In contrast, meteoritic analyses have consistently uncovered a diverse suite of larger S-bearing molecules, most of which contain more than nine atoms, for example, benzothiophene C$_8$H$_6$S \cite{Yabuta2007,orthous2010speciation,mojarro2023murchison}. The absence of these larger, potentially prebiotically relevant S species in the ISM raises critical questions about the chemical evolution that bridges interstellar matter and the chemical composition of cometesimals and planetesimals. It is possible that our current observational capabilities are insufficient to detect these elusive large compounds in the ISM or that substantial chemical transformations occur during the transfer from interstellar clouds to planetary surfaces, such as the early Earth.

To date, more than 340 interstellar molecules have been detected in the interstellar medium and/or circumstellar envelopes \cite{araki2025,mcguire20222021}. The record of these discoveries is characterized not merely by a gradual increase in numbers but also by the episodic emergence of entirely new molecular groups. The most recent breakthrough occurred following the detection of benzonitrile by McGuire et al. \cite{mcguire2018detection} in 2018 within the Taurus Molecular Cloud 1 (TMC–1). Subsequently, several cyclic hydrocarbon cyanides have been identified, with 1- and 2-cyanonaphthalene ($c$-C$_{10}$H$_7$CN) being recognized as the first polycyclic hydrocarbons \cite{mcguire2021detection}. Following this, tetracyclic hydrocarbon cyanides, such as 1-, 2-, and 4-cyanopyrene ($c$-C$_{16}$H$_9$CN) \cite{wenzel2024detection,wenzel2024detections} have been found in 2024. Currently, the largest interstellar molecule, apart from fullerenes, is the 7-ring polycyclic aromatic hydrocarbon (PAH) cyanocoronene ($c$-C$_{24}$H$_{11}$CN) \cite{wenzel2025discovery}.
It is important to note that the detected cyclic hydrocarbons include not only cyanide species but also pure hydrocarbons, for example, ethynylbenzene ($c$-C$_{6}$H$_5$CCH) \cite{loru2023detection}. Among these, indene ($c$-C$_9$H$_8$), a bicyclic hydrocarbon, is the largest species identified to date \cite{cernicharo2021pure}. In total, 17 cyclic hydrocarbon species, including benzene \cite{cernicharo2001infrared}, have been observed.
In a natural progression of these findings, Yang et al. \cite{yang2024have} have proposed that S-bearing PAHs, particularly those incorporating S heterocycles, might serve as a reservoir for the missing S. This proposal is supported by the detection of compounds such as thiophenol  ($c$-C$_6$H$_5$SH) \cite{Yabuta2007}, diphenyl disulfide, dibenzothiophene, thianthrene \cite{orthous2010speciation}, and thiophene along with its related species \cite{mojarro2023murchison} in meteorites. Nevertheless, although S-bearing species account for approximately 15\% of the interstellar molecules detected, S-bearing cyclic hydrocarbons have yet to be identified in interstellar space except for the claimed detection of the small species $c$-C$_3$H$_2$S \cite{remijan2025apj}, possibly due to their missing spectral experimental characterization. 

Although thiophenol represents one of the most fundamental S-bearing cyclic hydrocarbons and it is detected in meteoritic material, its dipole moment is relatively small (a- and b-dipole moments: $\mu_a$ = 0.83\,D and $\mu_b$ = 0.75\,D, see Extended Data Table 1), which might represent a major limiting factor for its possible identification in the interstellar gas through millimeter-wave astronomical observations. Also, the rotation of the SH-group relative to the phenyl ring causes a splitting of its rotational lines, and hence an overall spread of its emitted intensity (and therefore molecular population) across a higher number of rotational levels. In contrast, the 2,5- and 2,4-cyclohexadien-1-thione (two isomeric forms of $c$-C$_6$H$_6$S, hereafter 2,5-CT and 2,4-CT, respectively; a-type dipole moments: 4.73 and 3.87\,D, Extended Data Table 1), structural isomers of thiophenol, do not exhibit splitting due to spin-rotation interactions, internal rotation, or hyperfine effects. These rigid molecules display a pure rotational spectrum characteristic of an asymmetric top—analogous to the behavior observed in 2,5- and 2,4-cyclohexadien-1-one ($c$-C$_6$H$_6$O) \cite{mccarthy2020exhaustive}—and, as discussed later, possess substantially larger dipole moments. Consequently, the rotational lines of these two species might be brighter than those of their most stable isomeric form, rendering them favorable for interstellar detection in molecular clouds. Similar cyclic hydrocarbons incorporating a doubly hydrogenated carbon (-CH$_2$-) within their ring structures, have already been detected in interstellar gas (for example cyclopentadiene \cite{cernicharo2021discovery}, ethynyl cyclopentadiene \cite{cernicharo2021discovery}, indene \cite{cernicharo2021pure}, 1-cyanocyclopentadiene \cite{mccarthy2021interstellar}, and 2-cyanocyclopentadiene \cite{lee2021interstellar}), stressing the potential interest of these thiophenol isomers for interstellar detection.

We report here the discovery of 2,5-cyclohexadien-1-thione (2,5-CT), the largest S-bearing molecule detected so far, toward the Galactic center (GC) molecular cloud G+0.693-0.027 (hereafter, G+0.693). Situated within the Sgr B2 complex, this source is one of the principal reservoirs of complex organic molecules (COMs, organic molecules with six or more atoms in total) in our galaxy and has been the site of over 20 first interstellar detections (see, for example, \cite{rivilla2022a,jimenez-serra2022,Rivilla23,SanAndres2024}). We selected G+0.693 for our search for 2,5-CT due to its exceptional richness in S-bearing species \cite{rodriguez-almeida2021a,Rey-Montejo2024,Sanz-Novo2024a,Sanz-Novo2024b,sanz2025abiotic}, and because the aromatic ring benzonitrile has also been recently detected (Rivilla, V. M. priv. communication). This chemical diversity is likely attributable to enhanced sputtering of icy grain mantles induced by large-scale shocks from cloud–cloud collisions, which release a significant fraction of the S budget expected to be locked in interstellar ices \citep{zeng2018}. \\

\section{Results and discussion}

\subsection{Laboratory identification}

The experimental search of the 2,5-CT and 2,4-CT was guided by \textit{ab-initio} quantum chemical calculations to derive the main spectroscopic parameters from geometry optimization and harmonic force field analysis (see Section\,\ref{ch:theo_calc} and Extended Data Figure 1). The rotational transitions of 2,5-CT and 2,4-CT in the 8--40\,GHz range were observed using a combination of a chirped pulse Fourier Transform microwave spectrometer and a pulsed-discharge supersonic jet (see Section \,\ref{ch:exp} and Extended Data Figures 2 and 3 for details). Molecules were generated by a pulsed discharge in the throat of a 10-Hz supersonic jet using a vapor pressure of thiophenol at room temperature of 25 $^\circ$C. The experimental settings were optimized by monitoring the production of 2,5- and 2,4-cyclohexadien-1-one generated from the discharge of anisole ($c$-C$_6$H$_5$OCH$_3$). 

In the 8--40\,GHz region, 92 and 75 rotational lines of 2,5-CT and 2,4-CT, respectively, have been detected, covering quantum numbers up to $J$\,=\,15 and $K_a$\,=\,7, as shown in the Supplementary Data C6H6S-25.txt and C6H6S-24.txt \citep{piform2025}. Their rest frequencies, determined with a precision of about 5\,kHz, have been fitted to an effective Watson-type Hamiltonian in S-reduction including rotational and centrifugal distortion constants as listed in Table \ref{table:Parameters}. The three rotational constants $A_0$, $B_0$, and $C_0$ and the centrifugal distortion constants $D_J$, $D_{JK}$, $d_1$, and $d_2$ have been determined by the fit, while the centrifugal distortion constant $D_K$ was kept fixed to the values obtained by the CAM-B3LYP/cc-pCVTZ calculations \citep{yanai2004new,woon1995gaussian}. The fit reproduces the experimental frequencies with a root mean square deviation of 2.8 and 3.3\,kHz for 2,5-CT and 2,4-CT, respectively. Based on this analysis, rest frequencies for transitions used in the analysis of the astronomical spectra
are predicted with uncertainties sufficiently better than 10\,kHz. For the frequency ranges of the spectroscopic survey carried out toward the G+0.693 cloud, this uncertainty corresponds to uncertainties in velocity by less than 0.1 km s$^{-1}$. This is small compared to the typical linewidths of the molecular emission measured toward this cloud (of $\sim$20\,km\,s$^{-1}$ \citep{zeng2018}).

\subsection{Detection of 2,5-cyclohexadien-1-thione in G+0.693}

We analyzed an unbiased, ultrasensitive broadband spectral survey of G+0.693 carried out with the IRAM 30-m and Yebes 40-m radiotelescopes (for details of the observations see Methods Section). The observed data was compared with simulated spectra of 2,5-CT generated with the Spectral Line Identification and Modeling (SLIM) tool within the \textsc{Madcuba} software package \citep{martin2019}, under the assumption of constant excitation temperature, here referred to as local thermodynamic equilibrium (LTE). We note that the intermediate H$_2$ volume densities (10$^4$\,--\,10$^5$\,cm$^{-3}$) in G+0.693 \citep{zeng2018,Colzi2024} result in subthermal excitation of molecular emission, yielding excitation temperatures (i.e., $T_{\rm ex}$ = 5\,--\,20\,K, lower than the kinetic temperature of $T_{\rm k}$ = 50\,--\,150\,K \citep{zeng2018}). Unlike massive hot cores or low-mass hot corinos—where numerous rotational transitions, including those from vibrationally excited states, are observed—only low-energy rotational transitions in the ground vibrational state are detectable in G+0.693, significantly reducing the levels of line blending and confusion due to the excitation temperatures. Consequently, with the current sensitivity, we anticipate the detection of a few tens of transitions of 2,5-CT at these low excitation temperatures.

After assessing the emission of more than 140 molecules previously identified toward G+0.693, we detected a large number of $a$-type transitions of 2,5-CT with an integrated signal-to-noise (S/N) ratio $>$5 covering from the upper rotational levels $J$$_{\rm up}$\,=\,12 to $J$$_{\rm up}$\,=\,19, including several pairs of transitions belonging to two nearly complete progressions of ($J$+1)$_{0,J+1}$ ← $J$$_{0,J}$ and ($J$+1)$_{1,J+1}$ ← $J$$_{1,J}$ transitions (see Figure 1a), with the exception of the 12$_{1,12}$--11$_{1,11}$ transition, which fall out of the covered frequency range and the 17$_{1,17}$--16$_{1,16}$ and 17$_{0,17}$--16$_{0,16}$ transitions that appear heavily blended with unidentified (U) lines. These pairs of lines progressively converge as the frequency increases, eventually coalescing into a doubly degenerate line for $J$$_{\rm up}$\,=\,19. Overall, we found 22 unblended or slightly blended -contaminated by less than 25$\%$- features (see Figure 1, spectroscopic information listed in the Extended Data Table 2, including an analysis of the contamination), which were subsequently used to conduct the LTE fit and to derive the physical parameters of 2,5-CT (detailed information of the LTE fitting using the \textsc{Madcuba}-SLIM tool and the definition of unblended line is provided in the Methods Section). We stress that no missing lines are observed within the whole dataset, and the remaining lines are either heavily blended or too weak to be observed (i.e., transitions at 2\,mm and 3\,mm that do not arise above the noise), but are also in agreement with the observed spectra.

The best fitted LTE model for 2,5-CT (shown in red in Figure 1) yields an excitation temperature of $T$$_{\rm ex}$\,=\,14.3\,$\pm$\,3.4\,K, a radial velocity of $\varv$$_{\rm LSR}$\,=\,71.7\,$\pm$\,0.9\,km\,s$^{-1}$, a linewidth of FWHM = 20.0\,km\,s$^{-1}$ and a molecular column density of $N$\,=\,(5.6\,$\pm$\,0.3)\,$\times$10$^{12}$\,cm$^{-2}$, which translates into a fractional abundance with respect to H$_{2}$ of (4.1\,$\pm$\,0.7)\,$\times$\,10$^{-11}$, using $N$(H$_{2}$)\,=\,1.35\,$\times$\,10$^{23}$\,cm$^{-2}$ as derived by \cite{martin_tracing_2008}. A complementary population diagram analysis has also been performed \citep{goldsmith1999}, obtaining physical parameters that are in good agreement with the SLIM-\textsc{Autofit} analysis: $N$\,=\,(5.6\,$\pm$\,1.4)\,$\times$10$^{12}$\,cm$^{-2}$, and $T_{\rm ex}$\,=\,12.5\,$\pm$\,1.5\,K (see Methods Section and Extended Data Figure 4). The partition functions used are listed in the Extended Data Table 3.

\subsection{Astrophysical implications}

The detection of 2,5-CT opens a new window into the chemistry of large S-bearing cyclic molecules in the ISM, providing the first basis for elucidating their abundance and formation. The most straightforward comparison is between 2,5-CT and its structural isomers, 2,4-CT and thiophenol, which are not clearly detected in the current astronomical data (see Methods Section). Based on the derived upper limits for both molecules (i.e., $N$(2,4-CT) $\leq$\,3.2\,$\times$\,10$^{12}$\,cm$^{-2}$ and $N$(thiophenol) $\leq$\,8\,$\times$\,10$^{13}$\,cm$^{-2}$), we expect that 2,5-CT is a factor of $>$\,2 more abundant than 2,4-CT (which is similar to the factor of $\sim$\,2 estimated in our laboratory by relative intensities of rotational lines) while $N$(thiophenol)/$N$(2,5-CT) $<$\,14. Between the two structural isomers 2,4-CT and 2,5-CT, according to the energy level diagram (Extended Data Figure 1), it is expected that the low energy one is more abundant, in agreement with the detection, but the relatively low dipole moment of thiophenol compared to both 2,5-CT and 2,4-CT (see Extended Data Table 1) prevents us from unveiling conclusively whether only 2,5-CT is selectively produced in the ISM, or if 2,4-CT and thiophenol are also present but remain undetected due to sensitivity limitations. Besides the emergence of 2,5-CT as the only structural isomer identified to date for the C$_6$H$_6$S family, our findings now confirm the existence of large ($>$10 atoms) sulfur-containing cyclic species in the ISM. Although 2,5-CT itself accounts for only a small fraction of the s budget detected toward G+0.693 so far ($\sim$\,0.05\,$\%$; Sanz-Novo Private comm.), its discovery may be just the tip of the iceberg of a yet unexplored chemistry. This scenario might closely mirror that of benzonitrile, whose initial detection in the ISM by McGuire et al. \cite{mcguire2018detection} preceded the discovery of numerous cyano-substituted PAHs, and is in line with the rich inventory of S-bearing rings in meteorites, with over 80 species detected (including thiophenol, dibenzothiophene, and thianthrene \cite{Yabuta2007,orthous2010speciation}. By analogy with the nearly flat abundance trend of cyano-bearing rings found in TMC-1 \cite{wenzel2024detection,wenzel2025discovery}, the cumulative contribution of the C$_6$H$_6$S isomers in G+0.693 could approach $\sim$1.5$\%$ of the total S reservoir, hinting that S-bearing cyclic hydrocarbons and S-bearing PAHs might not represent a relevant sink of sulfur in the ISM. In this context, if a small portion of sulfur is locked up in S-bearing PAHs and related S-containing cyclic species, we anticipate that James Webb Space Telescope (JWST) observations would be capable of detecting several IR features, \cite{yang2024have} even though some of the prominent bands (for example, 10 $\mu\mathrm{m}$ C–S band) may be obscured by the 9.7 $\mu\mathrm{m}$ silicate absorption band. Meanwhile, upcoming radioastronomical facilities such as the Square Kilometre Array (SKA) or the Atacama Large-Aperture Submm/mm Telescope (AtLAST) will likely unravel a rich reservoir of large cyclic S-bearing species, including potentially prebiotic molecules.

The detection of 2,5-CT can be rationalized in terms of its large dipole moment (a-type dipole moment: $\mu_a$ = 4.73\,D, Extended Data Table 1), and establish this organosulfur species as a promising observational link between the rich S inventory found in the minor bodies of the solar system (i.e., asteroids, comets and meteorites), which includes a wide array of cyclic S-bearing compounds, ranging from thiophenol \cite{Yabuta2007} and thiophene \cite{mojarro2023murchison}, to the more complex diphenyl disulfide, dibenzothiophene and thianthrene \cite{orthous2010speciation}, and the known S-budget in the ISM, which has been limited so far to the detection of molecules up to 9 atoms \cite{kolesnikova2014spectroscopic,rodriguez-almeida2021a,sanz2025abiotic}. With 13 atoms, 2,5-CT now ranks as the largest S-bearing molecule detected so far in the ISM, marking a important step forward in molecular size and complexity within interstellar sulfur chemistry. Previously, the largest S-bearing interstellar species contained up to nine atoms (for example, ethyl mercaptan, CH$_3$CH$_2$SH, and its isomer dimethyl sulfide, CH$_3$SCH$_3$ \cite{rodriguez-almeida2021a,sanz2025abiotic}). In this context, 2,5-CT represents the largest S-bearing COM detected so far and also the most complex S-bearing cyclic species, providing a novel view on cyclic interstellar chemistry, which now extends beyond pure hydrocarbons and their cyano (-CN) and ethynyl (-CCH) derivatives. Additionally, our findings highlight the need for caution when analyzing mass spectrometric measurements of cometary, meteoritic, and asteroid material targeting thiophenol, as its mass peak could be contaminated by 2,5-CT given that both molecules share the same molecular mass (110.02\,u). Consequently, although 2,5-CT has not, to our knowledge, been searched for in extraterrestrial material from these minor bodies, it may still be present and yet may remain unidentified.

To date, the potential formation routes for 2,5-CT and related S-rings remain largely unexplored, both experimentally and theoretically. Consequently, we can only hypothesize its possible formation routes by either studying chemically related species or by drawing analogies with bottom-up pathways proposed for related cyclic species. Given the uncertain efficiency of gas-phase pathways, such as those invoked for the benzene formation through ion–molecule reactions \cite{Kocheril2025}, considered as the bottle neck in the growth of larger PAHs, a potential dust-grain origin appears particularly promising in G+0.693. Laboratory simulations have shown that cosmic-ray irradiation of low-temperature acetylene (C$_2$H$_2$) ices efficiently produces benzene \cite{Zhou2010}, suggesting that an analogous chemistry involving small S-bearing carbon chains (for example, C$_2$S, C$_3$S, detected in the ISM up to the 5-carbon member, C$_5$S \cite{Cernicharo21a}) and C$_2$H$_2$ on icy grains could lead to 2,5-CT. Although this hypothesis still needs to be tested in the laboratory, it is supported by two key factors that shape the chemistry of G+0.693, where linear S-chains are also abundant (for example, $N$(CCS)\,=\,1.5\,$\times$\,10$^{14}$\,cm$^{-2}$): i) Its elevated cosmic-ray ionization rate (10$^{-14}$--10$^{-15}$\,s$^{-1}$), estimated through chemical modelling involving cations such as PO$^+$ and HOCS$^+$ \cite{Sanz-Novo2024a} and references therein, which favours radical formation and recombination on grain mantles \cite{Rivilla23}. ii) The presence of large-scale low-velocity shocks, associated with a cloud–cloud collision scenario \cite{zeng2020}, which enhance the sputtering of icy grain mantles and could facilitate the desorption of molecules such as 2,5-CT. An analogous connection has already been suggested between benzonitrile, which is also detected in G+0.693 (Rivilla, V. M. priv. communication), and the cyanopolyyne family \cite{jose2021molecular}, but in the case of 2,5-CT, the inclusion of ring defects –i.e., a (-CH$_2$-) moiety that disrupts the electron delocalization, and thus the aromaticity, within the ring– needs to be addressed. Alternatively, benzene could be directly released from the grains through shocks and subsequently react through radical-neutral reactions \cite{lee2019gas,gagonova2024radiation}, which are considered among the main formation routes for diverse PAHs including: $c$-C$_6$H$_5$CN \citep{mcguire2018detection}, $c$- C$_{10}$H$_7$CN \cite{mcguire2021detection}, $c$-C$_5$H$_5$CN \cite{mccarthy2021interstellar}, and $c$-C$_{16}$H$_9$CN \cite{wenzel2024detection,wenzel2024detections}. However, apart from a recent study on the production of $c$-C$_3$H$_2$S via $c$-C$_3$H$_2$ + SH \cite{remijan2025apj}, there is lacking theoretical and experimental data that support an analogue formation route starting from $c$-C$_6$H$_6$ and yielding 2,5-CT or any of its isomers (for example, thiophenol).

In summary, the study of interstellar chemistry of large (i.e. $>$\,12 atoms) cyclic species has been bound so far to pure cyclic hydrocarbons or cyano (-CN) and ethynyl (-CCH) derivatives, as their derivatization provides a sizable dipole moment to the parental, typically nonpolar hydrocarbon (for example, benzene, naphtalene or pyrene) thus enabling their radioastronomical identification. The interstellar detection of 2,5-CT presented here, based on new high-resolution rotational data, demonstrates that interstellar cyclic chemistry extends beyond the aforementioned families to encompass S-bearing compounds. These findings open the window to a yet uncharted S-chemistry which, although it might not account for the missing sulfur in dense interstellar environments, does contribute considerably to our understanding of the origin of sulfur-containing molecules in meteorites and comets and provides insight into sulfur reservoirs in young Solar-type systems.

\section{Methods}\label{ch:methods}

\subsection{Theoretical calculations}\label{ch:theo_calc}

All the quantum chemical calculations were carried out with ORCA \citep{RN178} using CAM-B3LYP \citep{yanai2004new}. The correlation-consistent polarized core‐valence quintuple-zeta basis set (i.e., cc-pCV5Z) \citep{woon1995gaussian} was used for molecular geometry optimizations and energy calculations of thiophenol, 2,5-CT, and 2,4-CT. Effective rotational constants and centrifugal distortion constants for 2,5-CT and 2,4-CT were evaluated theoretically at cc-pCVTZ through harmonic force field calculations. Calculated energies and dipole moments are listed in the Extended Data Table 1; rotational and centrifugal distortion constants are shown in Table 1 along with their experimental values. The accuracy of dipole moments is expected to be around 8.5\% based on the comparisons of the calculated values by this method with the observed dipole moments for H$_2$S, dimethylsulfide, thiophene, and 3-methylthiophene.

\subsection{Laboratory measurements}\label{ch:exp}

High-resolution rotational spectra of 2,5-CT and 2,4-CT were observed using a high-resolution broadband microwave spectrometer in combination with a pulsed-discharge supersonic jet, as shown in the Extended Data Figures 2 and 3. 

\subsubsection{Pulsed-discharge supersonic jet}

The molecules of 2,5-CT and 2,4-CT were generated via pulsed discharge in a 10-Hz supersonic jet (CASJet, \citep{Lattanzi18}) using thiophenol (Thermo Fisher Scientific, without further purification) maintained at its vapor pressure at room temperature (25 $^\circ$C), diluted in neon as a buffer gas with a flow of 50–55\,sccm (standard cc/min). The jet was produced with a pulse valve (Parker) controlled by a pulse driver (IOTA ONE, Parker). Employing a backing pressure of $\sim$\,1\,kTorr, the molecular beam was cooled to a rotational temperature of approximately 5\,K, as estimated from the relative line intensities of thiophenol. Under these conditions, the pressure in the vacuum chamber was maintained at approximately 0.5$\times$10$^{-4}$ -- 1.0$\times$10$^{-4}$\,Torr using a combination of a diffusion pump, a root blower, and a rotary pump.

The discharge nozzle, mounted directly after the pulse valve, comprised electrodes made of small copper disks with a thickness of 3 mm, each featuring a central aperture through which the molecular beam flowed—4.3 mm for the downstream electrode and 2.3 mm for the upstream electrode. The downstream electrode was grounded, serving as the anode, while the upstream electrode was negatively charged, functioning as the cathode. These electrodes were separated by a Teflon insulator cylinder (13.4 mm in length, with a 3.5 mm central aperture) and connected in series with a 50\,k$\Omega$ ballast resistor. A voltage of 1000\,V was applied across the assembly, resulting in a current of a few milliamperes. Under these conditions, a plasma discharge was generated between the electrodes immediately prior to the supersonic expansion in the vacuum chamber.

\subsubsection{The chirped pulse Fourier Transform microwave spectrometer}

The chirped pulse Fourier Transform spectrometer is a high resolution broadband instrument covering frequencies in the range 8--40\,GHz. An intense pulse of 2~\textmu s length, chirped in frequency, produces a macroscopic polarisation of the molecular sample. The subsequent free induction decay (FID) is recorded in the time domain using a heterodyne receiver. 
A two-channel arbitrary waveform generator (Keysight, M8190A) generates the chirped pulse, which is frequency up-converted using an IQ modulator and a tunable signal generator (Agilent Technologies, E8257D) to cover the frequencies of interest (8--18\,GHz, 18--26\,GHz, 26--40\,GHz). A solid state amplifier (8--18\,GHz: Microsemi C0618-43-T680,  18--26\,GHz: Microsemi C1826-36-T964, 26--40\,GHz: Eravant SBP-2734033530-KFKF-S1-HR) amplifies the signal before it is emitted via a quad ridge horn antenna, the other port of which is also used to detect the molecular signal. A roof top mirror in the chamber is used to rotate the polarisation by 90$^\circ$.
The receiver consists of a low noise amplifier protected from the intense excitation pulse by a fast pin diode switch, followed by an identical IQ modulator using the same local oscillator signal to down convert the FID. The intermediate frequency (IF) signal is then fed into a 5~GHz low pass filter, followed by another amplifier, before being digitised using an Acqiris U9510A digitiser card. The measurements are repeated and averaged in the time domain to improve the signal-to-noise ratio. During the measurement, the phase of the chirped pulse is cycled (0~$^{\circ}$, 90~$^{\circ}$, 180~$^{\circ}$, 270~$^{\circ}$), which allows sideband separation and suppression of spurious harmonics.

\subsection{Astronomical observations}
\label{Observations}

We have analyzed a new unbiased, ultradeep molecular line survey conducted toward the Galactic Center molecular cloud G+0.693, using the Yebes 40-m (Guadalajara, Spain) and the IRAM 30-m (Granada, Spain) radiotelescopes. This broadband survey spans a frequency range of $\sim$91\,GHz and exhibits an increased sensitivity compared the data used in previous works (for example, \citep{rodriguez-almeida2021a}). The observations were carried out using the position switching mode toward the equatorial coordinates of G+0.693 (i.e., $\alpha$ = $\,$17$^{\rm h}$47$^{\rm m}$22$^{\rm s}$, $\delta$ = $\,-$28$^{\circ}$21$^{\prime}$27$^{\prime\prime}$), using an off position shifted by $\Delta\alpha$~=~$-885$$^{\prime\prime}$ and $\Delta\delta$~=~$290$$^{\prime\prime}$. 

\subsubsection{Yebes 40-m radiotelescope}

New Yebes 40-m observing runs (project 21A014; PI: V. M. Rivilla) were performed between March 2021 and March 2022. We used the ultra broadband Nanocosmos Q-band (7\,mm) high electron mobility transistor (HEMT) receiver, which enables broadband observations across the whole $Q$-band (i.e., 18.5\,GHz between 31.07 and 50.42\,GHz) in two linear polarizations \citep{tercero2021}. The 16 fast Fourier transform spectrometers (FFTS) provided a raw channel width of 38\,kHz. We used two distinct spectral setups, centred at 41.4 and 42.3\,GHz, to identify possible spurious lines. The detailed procedure employed for the data reduction, combination and averaging of both the new Yebes 40-m and the IRAM 30-m data is presented in \citep{Rivilla23}. Subsequently, the spectra was imported into \textsc{Madcuba} \citep{martin2019} smoothed to a frequency resolution of 256\,kHz (i.e., velocity resolutions of 1.5$-$2.5\,km s$^{-1}$ in the range observed). An extraordinary sensitivity has been reached, with rms noise levels ranging between 0.25$-$0.9 mK across the whole $Q$-band at this spectral resolution in units of antenna temperature ($T$$\mathrm{_A^*}$) scale, since the molecular emission toward G+0.693 is extended over the beam \citep{Zheng2024}. The half power beam width (HPBW) of the telescope ranged between $\sim$35$-$55$^{\prime\prime}$ (at 50 and 31\,GHz, respectively). 

\subsubsection{IRAM 30-m radiotelescope}

The new IRAM 30-m observations (project 123-22; PI: Jim\'enez-Serra) were carried out between February 1$-$18 2023. We employed the multi-band mm-wave receiver Eight MIxer Receiver (EMIR) and various frequency setups to cover three frequency windows between 83.2$-$115.41, 132.28$-$140.39, and 142$-$173.81\,GHz. Each frequency set up was shifted in frequency in order to identify possible contamination of spurious lines coming from the image band. We achieved an initial spectral resolution of 195\,KHz by using the Fast Fourier Transform Spectrometer (FTS200), even though we finally smoothed the spectra within \textsc{Madcuba} to 615\,kHz, which translates to velocity resolutions of 1.0$-$2.2 km s$^{-1}$ in the observed frequency range.  The half power beam width (HPBW) of the telescope varies between 14$-$29$^{\prime\prime}$ across the frequency range covered. We note that for those frequency ranges that are not covered within these new data, we used the previous IRAM 30-m survey (further details are given elsewhere; for example, \citep{rodriguez-almeida2021a}). Overall, we obtained final noise levels between 0.5$-$2.5 mK at 3 mm, and 1.0$-$1.6 mK at 2 mm per channel.

\subsection{LTE Analysis of  2,5-CT with \textsc{Madcuba}}

Once the rotational spectroscopic data of 2,5-CT was imported into the \textsc{Madcuba} software package \citep{martin2019}, we used the Spectral Line Identification and Modeling (SLIM) tool (version from 2024, June 15) within \textsc{Madcuba} to analyze the astronomical data under the assumption of Local Thermodynamic Equilibrium (LTE). We generated the LTE simulated spectra with SLIM and then conducted a nonlinear least-squares LTE fit of the brightest transitions of 2,5-CT that are either unblended or exhibit a slight blending (shown in Figure 1 and listed in the Extended Data Table 2) to the observed spectra using the \textsc{Autofit} tool within SLIM \citep{martin2019}. The selection of these lines follows the criteria established in previous works \citep{Rey-Montejo2024,sanz2025abiotic} to identify unblended lines while considering potential contamination from a yet unidentified (U) line. To evaluate the level of line blending, we analyzed the region surrounding the 2,5-CT lines within a velocity range of $\pm$\,FWHM/2, where FWHM represents the observed linewidth. The level of contamination was assessed by subtracting the LTE fit of the 2,5-CT from the observed spectrum and calculating the residual area. The contribution to the residuals for all the selected lines are shown in the Extended Data Table 2. A line is classified as unblended if the residual area contributes 25$\%$ or less to the total. Additionally, if a known molecule that lies within this range contributes less than  25$\%$  of the total integrated intensity, the line will be considered to be slightly blended. The SLIM-\textsc{Autofit} method enables us to derive the following physical parameters: molecular column density ($N$), excitation temperature ($T_{\rm ex}$), radial velocity ($\varv$$_{\rm LSR}$) and FWHM. Only the latter parameter was fixed in the fit to a value of 20 km s$^{-1}$ to achieve convergence, which is in agreement with the characteristic FWHM measured for the molecular transitions in G+0.693 (FWHM $\sim$ 15$-$20 km s$^{-1}$; see for example, \citep{zeng2018}). We thus derived the following parameters: $N$ = (5.6 $\pm$ 0.3) $\times$ 10$^{12}$ cm$^{-2}$, $T_{\rm ex}$ = 14.3 $\pm$ 3.4\,K, and $\varv$$_{\rm LSR}$ = 71.7 $\pm$ 0.9 km s$^{-1}$. 

\subsection{Rotational diagram analysis}

As an alternative to the SLIM-\textsc{Autofit} analysis, we can profit from the rotational diagram method \citep{goldsmith1999} to derive the physical parameters of 2,5-CT. We have used the reduced subset of clean and slightly blended transitions listed in the Extended Data Table 2, excluding the transitions for which the contamination by a U-line accounts for $>$ 25\% of the overall area, We obtained results that are in good agreement with both the column density and excitation temperature obtained using SLIM-\textsc{Autofit}: $N$ = (5.6 $\pm$ 1.4) $\times$10$^{12}$ cm$^{-2}$, and $T_{\rm ex}$ = 12.5 $\pm$ 1.5\,K. The results of the rotational diagram are shown in the Extended Data Figure 4.

\subsection{Non-detection of 2,4-CT and thiophenol}

Regarding 2,4-CT, we also implemented its newly measured rotational data into \textsc{Madcuba}-SLIM and searched for it toward the survey of G+0.693. 2,4-CT is not clearly detected, despite the emergence of several weak and unblended transitions within the noise. However, unlike 2,5-CT, it lacks clear enough spectroscopic features in the astronomical data for a conclusive detection. Therefore, we used the LTE parameters obtained for 2,5-CT to derive the upper limit to its molecular abundance. We searched for the brightest predicted spectral features of 2,4-CT that appear to be completely unblended with emission from other molecules previously identified in the astronomical data. Specifically, we used the  13$_{0,13}$ -- 12$_{0,12}$ transition (located at $\sim$33.677\,GHz), which fall in one of the frequency regions of the survey with highest sensitivity (rms $\sim$0.5 mK), which enable us to place stringent constraints on the abundance of 2,4-CT toward G+0.693. We thus derived a 3$\sigma$ upper limit to its column density ($\sigma$ is the rms noise of the spectra) of $N \leq$\,3.2\,$\times$\,10$^{12}$\,cm$^{-2}$, which yields to an upper limit to the molecular abundance with respect to molecular hydrogen of 2.7\,$\times$\,10$^{-11}$ and does not produce any overly bright features at other frequencies. Based on the derived upper limit, 2,4-CT is a factor of $\sim$2 less abundant than 2,5-CT.

We have also searched for thiophenol using available laboratory rotational data \citep{Larsen2009}, without yielding a detection. Therefore, we derived the 3$\sigma$ upper limit to its molecular abundance using the 15$_{1,14}$ -- 14$_{1,13}$ transition (placed at $\sim$40.906\,GHz), which is the brightest and fully unblended transition predicted in the LTE model. We obtain a $N$ $\leq$ 8 $\times$ 10$^{13}$ cm$^{-2}$ adopting the physical parameters found for 2,5-CT. In terms of the molecular abundance with respect to H$_2$, the above value is translated into an upper limit of 6\,$\times$\,10$^{-10}$. We find that thiophenol is $\leq$\,14 times more abundant than 2,5-CT, in line with the energy ordering of the three structural isomers (see Extended Data Figure 1).

\subsection{Data availability statement}
This paper makes use of data from projects 018-19, 123-22 and 076-23 (IRAM 30-m), and 21A014 (Yebes 40-m). The observed spectra and fits of the transitions of the different species presented in this work are deposited in the Zenodo repository \cite{araki2026}.

\subsection{Code availability statement}
The MADCUBA package, which was used to perform the LTE analysis performed in this work, is a software publicly available at https://cab.inta-csic.es/madcuba/download.html. A description of the package is provided in \citep{martin2019}.

\section{Acknowledgment}
M.A., C.P.E., V.L., and P.C. acknowledge the Max Planck Society for the financial support. This paper makes use of data from projects 018-19, 123-22 and 076-23 (IRAM 30-m), and 21A014 (Yebes 40-m). 
The 40$\,$m radio telescope at Yebes Observatory is operated by the Spanish Geographic Institute (IGN, Ministerio de Transportes, Movilidad y Agenda Urbana). IRAM is supported by INSU/CNRS (France), MPG (Germany) and IGN (Spain). M.S.-N. acknowledges a Juan de la Cierva Postdoctoral Fellow proyect JDC2022-048934-I, funded by the Spanish Ministry of Science, Innovation and Universities/State Agency of Research MICIU/AEI/10.13039/501100011033 and by the European Union “NextGenerationEU”/PRTR”. V.M.R. acknowledges support from the grant RYC2020-029387-I funded by MICIU/AEI/10.13039/501100011033 and by “ESF, Investing in your future", from the Consejo Superior de Investigaciones Cient{\'i}ficas (CSIC) and the Centro de Astrobiolog{\'i}a (CAB) through the project 20225AT015 (Proyectos intramurales especiales del CSIC), and from the grant CNS2023-144464 funded by MICIU/AEI/10.13039/501100011033 and by “European Union NextGenerationEU/PRTR”.  D.S.A. also acknowledges support from the grant CNS2023-144464. I.J.-S., V.M.R., M.S.-N., L.C., A.M., A.L.-G., A.M.H., and D.S.A. acknowledge funding from grant No. PID2022-136814NB-I00 from MICIU/AEI/10.13039/501100011033 and by “ERDF, UE A way of making Europe”. I.J.-S. also acknowledges funding from the ERC grant OPENS (project number 101125858) funded by the European Union. Views and opinions expressed are however those of the author(s) only and do not necessarily reflect those of the European Union or the European Research Council Executive Agency. Neither the European Union nor the granting authority can be held responsible for them. M.S.-N., I.J.-S., L.C., and S.Z. acknowledge funding from Consejo Superior de Investigaciones Científicas (CSIC) through project i-LINK23017 SENTINEL. D.S.A. expresses his gratitude for the funds from the Comunidad de Madrid through Grant PIPF-2022/TEC-25475, and the financial support by the Consejo Superior de Investigaciones Cient{\'i}ficas (CSIC) and the Centro de Astrobiolog{\'i}a (CAB) through project 20225AT015 (Proyectos intramurales especiales del CSIC). S.Z. acknowledges the support by RIKEN Special Postdoctoral Researchers Program. J.G.dlC. acknowledges support from the grant No. PID2022-136814NB-I00 287 from MICIU/AEI/10.13039/501100011033 and by “ERDF, UE A way of making Europe”. J.G.dlC. also acknowledges European Funds of Regional Development, and the Autonomous Government of Extremadura (Grant GR24020).

\section{Author contributions statement}

M.A. and C.P.E. performed the laboratory experiments and collected the data; M.A. and V.L. carried out the theoretical calculations; C.P.E. analyzed the laboratory data. V.L. and P.C. coordinated the project. M.S.-N., V.M.R., I.J.-S., L.C., S.Z., A.M., A.L.-G., A.M.-H., D.S.A., S.M., M.A.R.-T., and J.G.dlC. contributed to the collection and reduction of the astronomical data. V.M.R. and I.J.-S. led the observational survey. M.S.-N. analyzed the astronomical observations. M.A. and M.S.-N. wrote the manuscript with the help of V.L. All authors provided feedback and commented on the manuscript.

\section{Competing interests statement}
The authors declare no competing interests.

\begin{table}[htbp]
\caption{Molecular parameters for 2,5- and 2,4-cyclohexadien-1-thione derived from the least square fits.}
\label{table:Parameters}
    { 
    \begin{tabular}{l l S[table-format=5.9] S[table-format=5.9] S[table-format=5.9] S[table-format=5.9]}
    \toprule
    \multicolumn{2}{l}{Parameter}
                       & \multicolumn{2}{c}{2,5-CT} 
                       & \multicolumn{2}{c}{2,4-CT} \\
    \cmidrule(r){3-4}\cmidrule(r){5-6}
     & & obs & calc & obs & calc \\
    \midrule
    $A_0$         & MHz &  5285.4602(38) &   5309.7 &   5216.7958(46)  &   5271.8 \\
    $B_0$         & MHz &  1608.45219(13)&   1618.3 &   1619.66382(15) &   1623.9  \\
    $C_0$         & MHz &  1242.73075(11)&   1249.7 &   1247.58786(13) &   1253.1  \\
    $D_J$         & kHz &    0.05122(19) &    0.049 &   0.06686(24)    &   0.054 \\
    $D_{JK}$      & kHz &    0.2562(24)  &    0.260 &   0.1806(70)     &   0.272 \\
    $D_K$         & kHz &    0.977377    & 0.977377 &   0.908029       &   0.908029 \\
    $d_1$         & kHz &   -0.01438(19) &   -0.014 &   -0.01527(22)   &   -0.015 \\
    $d_2$         & kHz &   -0.00281(14) &   -0.003 &   -0.00240(20)   &   -0.0022 \\
    \# lines      &     &    92         &        &  75            & \\
    RMS     & kHz &   2.8        &       & 3.3            & \\
    \bottomrule
    \bottomrule
    \end{tabular}}
    \begin{flushleft}
       {Note. 
       Spectroscopic constants of the $S$ reduced Hamiltonian in the I$^r$ representation were fitted using the program SPFIT
       \citep{pickett1991fitting}, and uncertainties of obtained spectroscopic parameters were evaluated by the program PIFORM \citep{piform2025}. Standard uncertainties are given in parentheses. Parameters without uncertainties have been kept fixed to the values obtained by the quantum chemical calculation. Calculated rotational constants were derived from the equilibrium values ($B_e$) at the cc-pCV5Z level with the vibrational corrections ($B_e - B_0$) estimated using cc-pCVTZ. Centrifugal distortion constants are calculated at cc-pCVTZ. RMS stands for root-mean-square error.}
\end{flushleft}
\end{table}
\FloatBarrier

   \begin{figure*}[htbp]
   \centering
   \includegraphics[width=1.0\columnwidth]{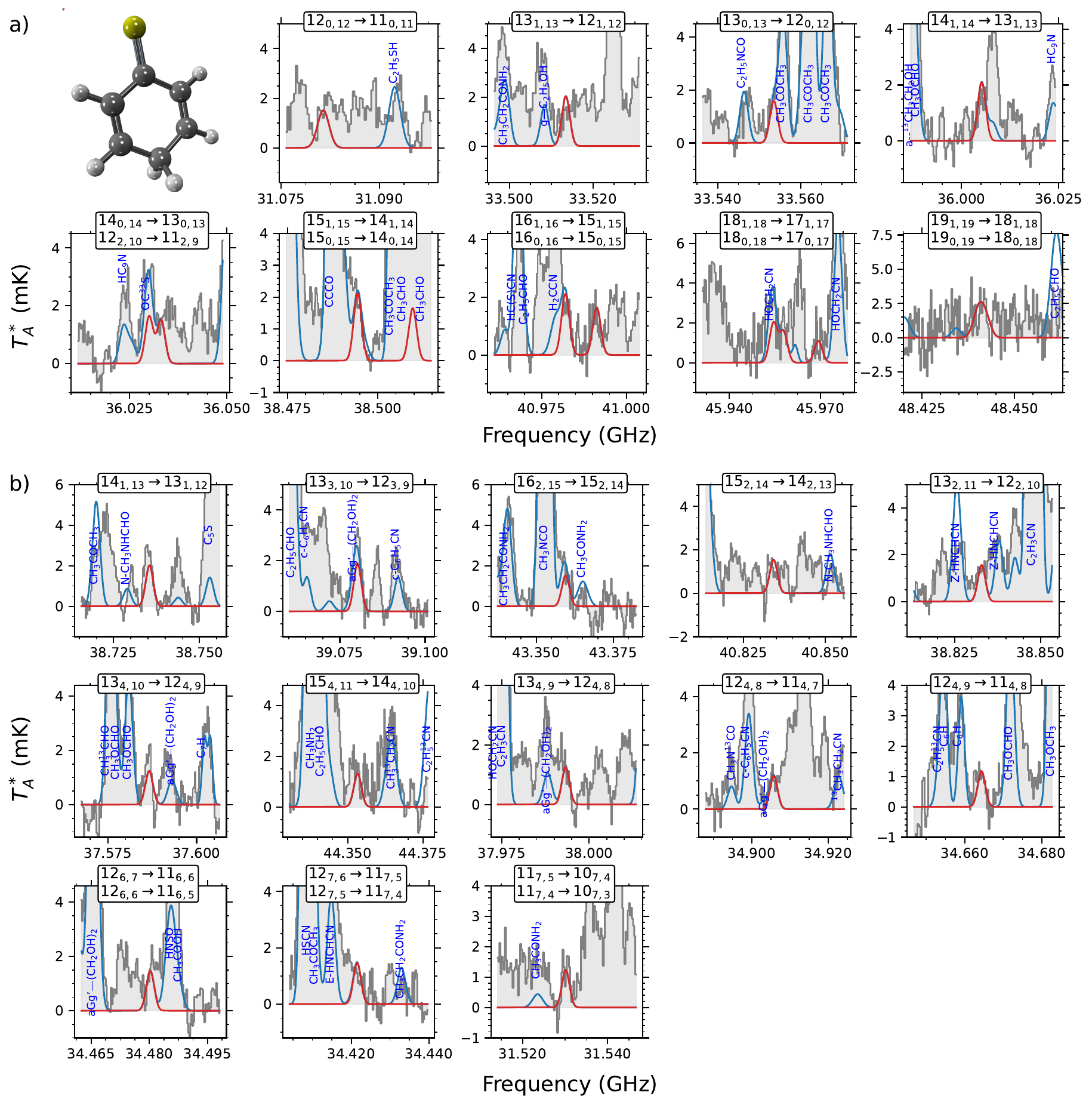}
   \caption*{\textbf{Fig. 1.} Detected lines of 2,5-CT toward the GC molecular cloud G+0.693–0.027. a) Pairs of $K_a = 0$ and 1 transitions 
   that progressively converge with increasing frequency, ultimately coalescing into a doubly degenerate line. b) $K_a>1$ transitions of 2,5-CT observed in the astronomical data that were also used to derive the LTE physical parameters of the molecule (see text; listed in the Extended Data Table 2). The quantum numbers involved in each transition are shown in the upper part of each panel. The red line depicts the result of the best LTE fit to the 2,5-CT rotational transitions, while the blue lines present the emission from all the molecules identified to date in the survey, including 2,5-CT, overlayed with the observed spectra (gray histograms and light gray shaded area). The 3D structure of 2,5-CT is also shown (carbon atoms in gray, S atom in yellow and hydrogen atoms in white).}
   \label{Fig:LTEspectrum}%
   \end{figure*}
\FloatBarrier


\newpage

\begin{figure*}
\captionsetup{labelformat=empty}
\centering
   \includegraphics[width=.95\columnwidth]{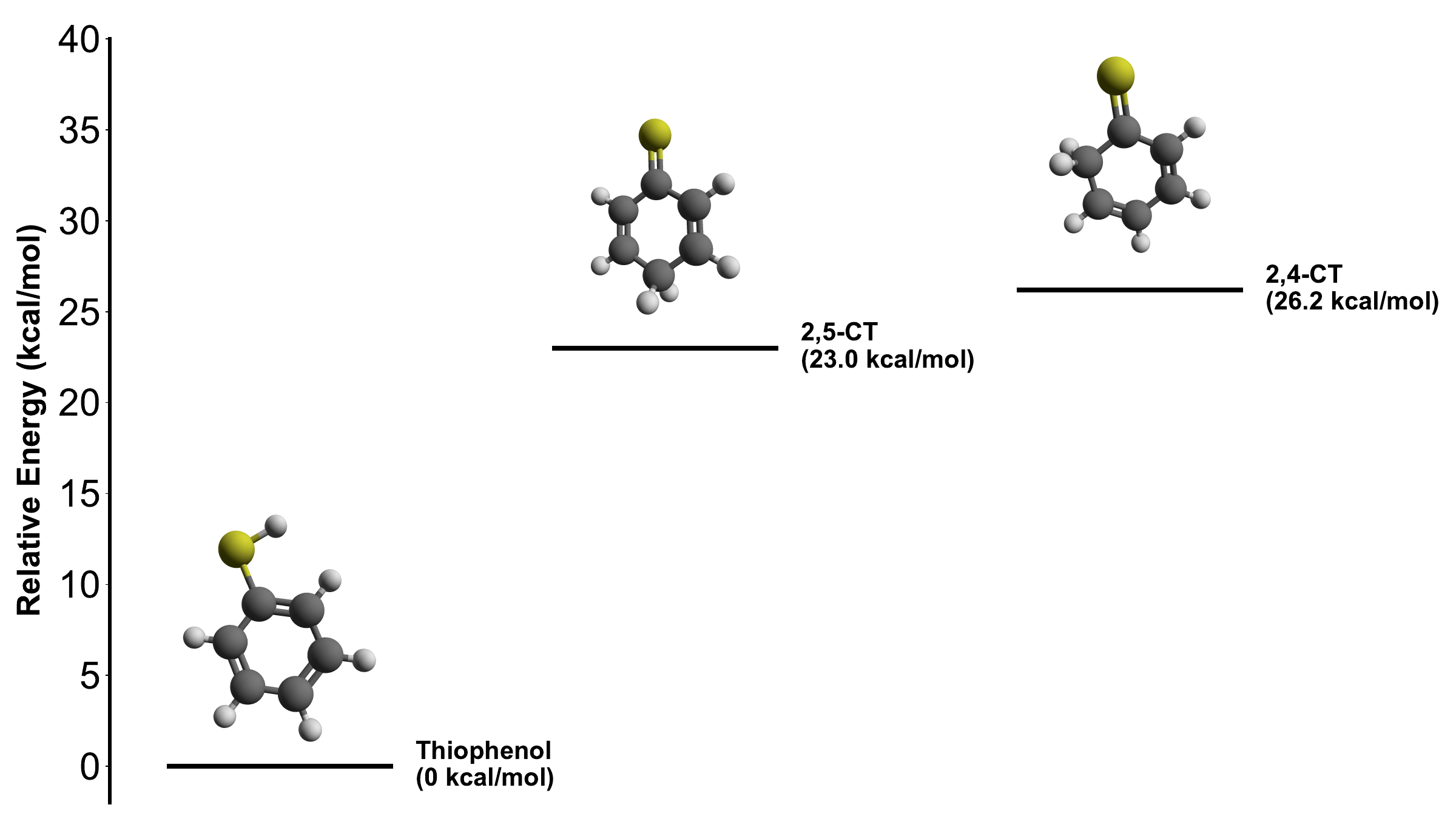}
\caption{Extended Data Figure 1.} Energy levels of 2,5-CT and 2,4-CT relative to thiophenol. The geometries and energies were derived at CAM-B3LYP/cc-pCV5Z level of theory.
\label{Fig:energies}%
\end{figure*}

\begin{figure*}
\captionsetup{labelformat=empty}
\centering
\includegraphics[width=.95\columnwidth]{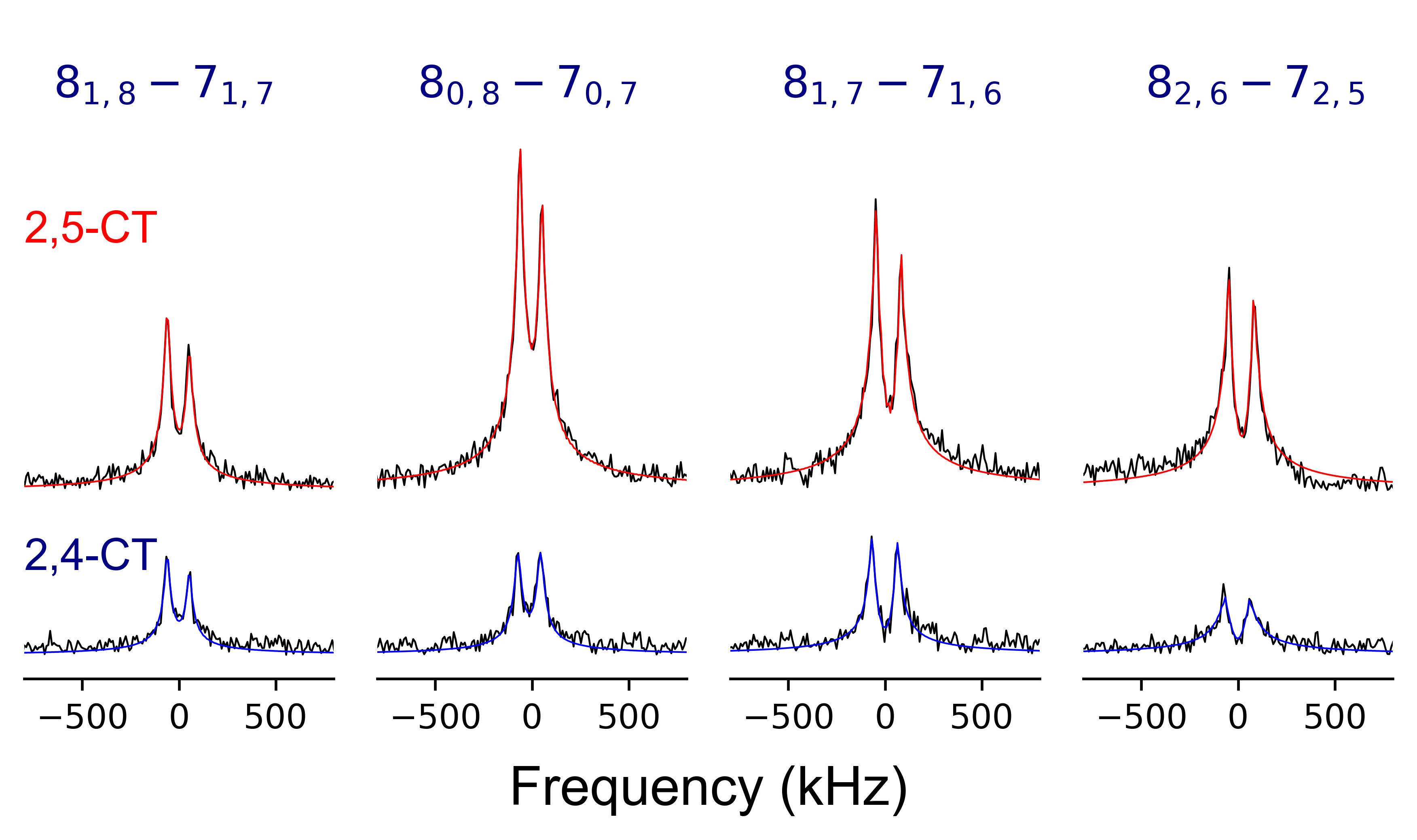}
\caption{Extended Data Figure 2. Sections of the experimental spectrum (black), showing four transitions for each of the two molecules 2,4-CT and 2,5-CT with corresponding quantum numbers $J'_{Ka',Kc'} - J''_{Ka'', Kc''}$ given in the top trace. The adapted line profile fit is shown in color. The frequency axis is centered on the predicted transition frequency in each case. The Doppler effect causes a double peak profile with a splitting of about 100 kHz due to the optical alignment.}
\label{Fig:Spectrum1}%
\end{figure*}

\begin{figure*}
\captionsetup{labelformat=empty}
\centering
\includegraphics[width=.95\columnwidth]{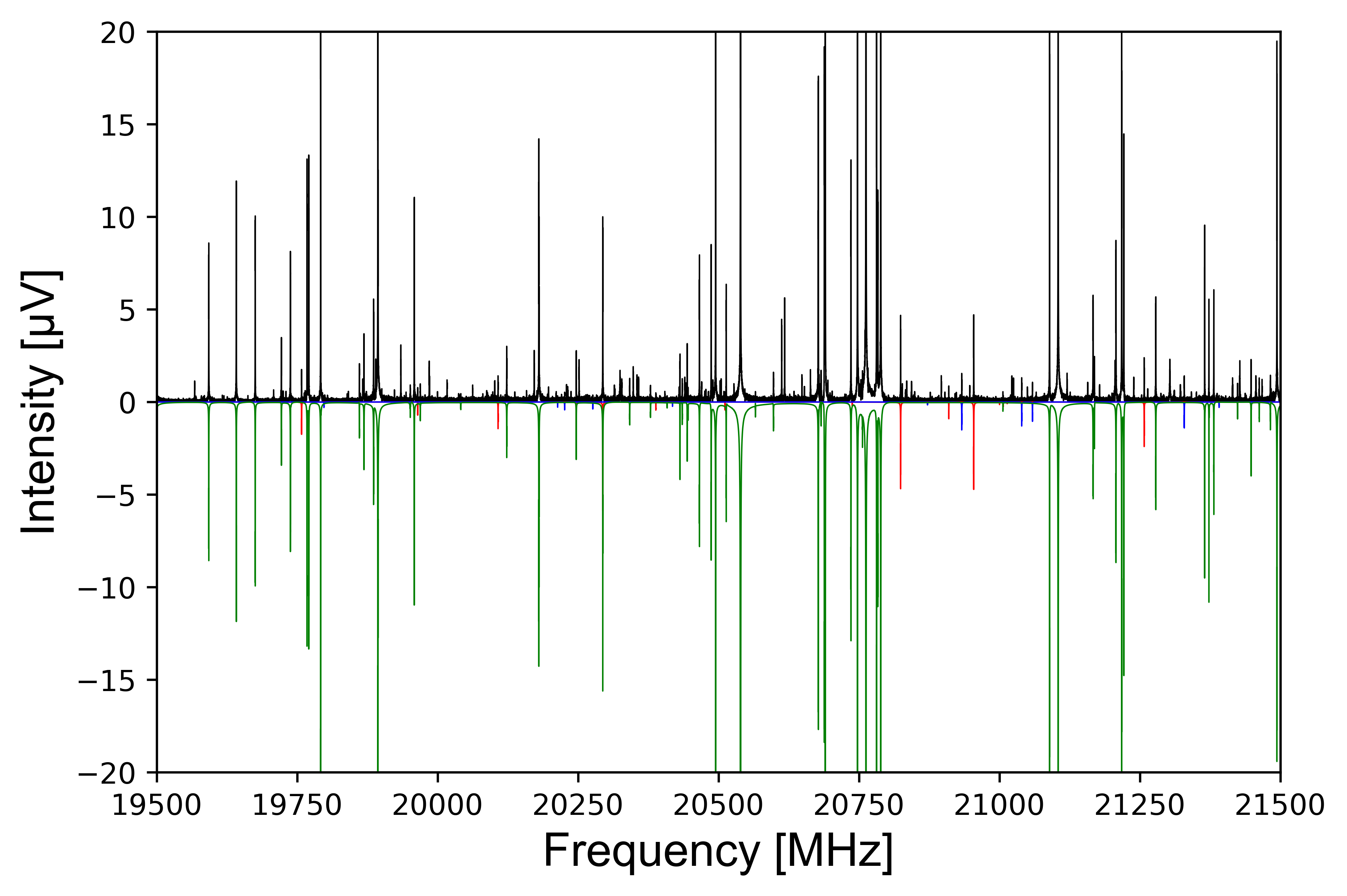}
\caption{Extended Data Figure 3. Section of the experimental spectrum (black, upward), showing the frequency range between 19500--21500\,MHz. The adapted line profile fit for selected species is shown in color (green: Thiophenol, black: $^{34}$S-Thiophenol, red: 2,5-CT, blue: 2,4-CT) with inverted intensities.}
\label{Fig:Spectrum2}%
\end{figure*}

\begin{figure}
\captionsetup{labelformat=empty}
\centerline{\resizebox{0.7\hsize}{!}{\includegraphics[angle=0]{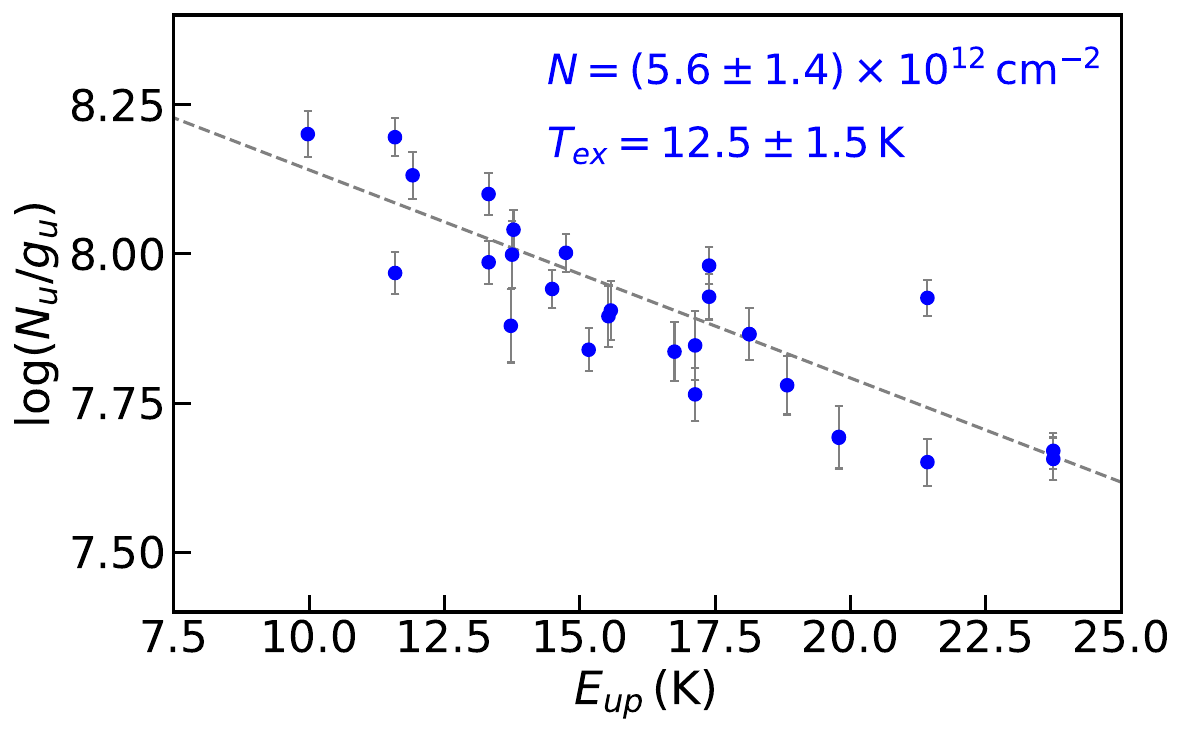}}}
\caption{Extended Data Figure 4. Rotational diagram for the selected transitions of 2,5-CT (depicted in Figure 1) observed toward G+0.693. Data points (blue dots) are presented as mean values $\pm$ standard errors of the means (1$\sigma$ errors). The best linear fit to the data points is shown using a gray dashed line. The values for the molecular column density, $N$, and the excitation temperature, $T_{\rm ex}$, obtained from the fit are shown in blue.} 
\label{f:rotdiagram}
\end{figure}

\begin{table*}
\centering                          
\captionsetup{labelformat=empty}
\caption{Extended Data Table 1. Energies and dipole moments for the thiophenol system. All the calculations were performed with ORCA using the CAM-B3LYP method and the cc-pCV5Z basis set.}             
\label{table:1}     
\centering                          
\begin{tabular}{c c c c c}      
\toprule
Species & Energy & Rel. En. & $\mu_a$ & $\mu_b$\\ 
 & [a.u.] & [kcal/mol] & [D] & [D] \\
\midrule
Thiophenol & -630.4423497 & 0.0 & 0.83 & 0.75\\
2,5-CT & -630.4056913 & 23.0 & 4.73 & \\
2,4-CT & -630.4005289 & 26.2 & 3.87 & 0.44\\
\bottomrule
\bottomrule
\end{tabular}{}
\end{table*}

\begin{sidewaystable}
\captionsetup{labelformat=empty}
\caption{Extended Data Table 2. Spectroscopic information of the brightest and cleanest detected transitions of 2,5-CT toward G+0.693$-$0.027, which are shown in Figure 1 and were used to derive the physical parameters of the molecule.}
\begin{tabular*}{\textheight}{@{\extracolsep\fill}ccccccccc}
\toprule%
Frequency & Transition$^{(a)}$ & log \textit{I} (300 K) & $E$$\mathrm{_{up}}$ &  rms & $\int$ $T$$\mathrm{_A^*}$d$\varv$ & Residual area $^{(b)}$ & S/N $^{(c)}$ & Blending \\ 
(GHz) &                        &  (nm$^2$ MHz)          &  (K)                &  (mK)& (mK km s$^{-1}$)              & (\%)  &          &    \\
\midrule
31.0816996 (13) & 12$_{0,12}$--11$_{0,11}$ & --5.2700 &  9.9 & 0.6 & 32.2 & 24.2 & 10.4 & Unblended \\   
31.5304333 (33) & 11$_{7,5}$--10$_{7,4}$   & --5.4181 & 18.0 & 0.6 & 27.0 & 24.4 & 8.7  & Unblended$^{*}$ \\ 
31.5304364 (33) & 11$_{7,4}$--10$_{7,3}$   & --5.4181 & 18.0 & 0.6 &      &      &      & Unblended$^{*}$ \\ 
33.5137909 (15) & 13$_{1,13}$--12$_{1,12}$ & --5.0626 & 11.5 & 0.5 & 26.3 & 5.7 & 10.7  & Unblended \\ 
33.5538160 (15) & 13$_{0,13}$--12$_{0,12}$ & --5.1707 & 11.5 & 0.5 & 40.3 & 7.0 & 16.4  & Slightly blended: CH$_3$COCH$_3$ \\
34.4218320 (35) & 12$_{7,6}$--11$_{7,5}$   & --5.2616 & 19.6 & 0.5 & 20.4 & 8.5 & 7.7   & Unblended$^{*}$ \\     
34.4218433 (35) & 12$_{7,5}$--11$_{7,4}$   & --5.2616 & 19.6 & 0.5 &      &     &       & Unblended$^{*}$ \\ 
34.4802711 (23) & 12$_{6,7}$--11$_{6,6}$   & --5.3101 & 17.3 & 0.5 & 24.9 & 19.3 & 9.3  & Unblended$^{*}$ \\     
34.4808071 (23) & 12$_{6,6}$--11$_{6,5}$   & --5.3101 & 17.3 & 0.5 &    &    &   & Unblended$^{*}$ \\ 
34.6646714 (10) & 12$_{4,9}$--11$_{4,8}$   & --5.2266 & 13.6 & 0.5 & 20.1 & 9.5 & 7.5 & Unblended \\  
34.9060892 (13) & 12$_{4,8}$--11$_{4,7}$   & --5.2206 & 13.7 & 0.5 & 20.2 & 7.3 & 7.6 & Unblended \\ 
36.0055197 (19) & 14$_{1,14}$--13$_{1,13}$ & --4.9702 & 13.2 & 0.4 & 27.4 & 3.0 & 12.1 & Unblended \\       
36.0303689 (19) & 14$_{0,14}$--13$_{0,13}$ & --5.0787 & 13.2 & 0.4 & 61.6 & 11.0 & 27.2 & Blended: OC$^{33}$S \\      
36.0334115 (18) & 12$_{2,10}$--11$_{2,9}$  & --5.1522 & 11.8 & 0.4 & 30.3 & 6.8 & 13.3 & Unblended \\
37.5869628 (14) & 13$_{4,10}$--12$_{4,9}$  & --5.1162 & 15.4 & 0.4 & 28.8 & 32.9 & 12.7 &  Blended: U-line \\ 
37.9936987 (20) & 13$_{4,9}$--12$_{4,8}$   & --5.1068 & 15.5 & 0.4 & 24.5 & 22.7 & 10.7 & Unblended \\ 
38.4947436 (26) & 15$_{1,15}$--14$_{1,14}$ & --4.8845 & 15.1 & 0.3 & 31.4 & 15.1 & 26.5 & Slightly blended: U-line \\
38.7357999 (21) & 14$_{1,13}$--13$_{1,12}$ & --4.9158 & 14.6 & 0.3 & 47.6 & 31.4 & 15.0 & Blended: U-line \\ 
38.8332164 (22) & 13$_{2,11}$--12$_{2,10}$ & --5.0545 & 13.7 & 0.3 & 28.0 & 13.4 & 19.3 & Unblended \\
39.0805294 (27) & 13$_{3,10}$--12$_{3,9}$  & --4.9508 & 14.4 & 0.3 & 50.5 & 15.6 & 7.9 & Slightly blended: $aGg'$-(CH$_2$OH)$_2$ \\ 
40.8344834 (28) & 15$_{2,14}$--14$_{2,13}$ & --4.9522 & 16.6 & 0.5 & 22.2 & 10.4 & 8.4 & Unblended \\  
40.9823609 (35) & 16$_{1,16}$--15$_{1,15}$ & --4.8046 & 17.0 & 0.5 & 39.3 & 7.1  & 14.9 &  Unblended \\         
40.9915911 (35) & 16$_{0,16}$--15$_{0,15}$ & --4.9135 & 17.0 & 0.4 & 24.0 & 11.2 & 10.1 &  Unblended \\
43.3598980 (36) & 16$_{2,15}$--15$_{2,14}$ & --4.8742 & 18.7 & 0.4 & 36.6 & 9.5  & 15.3 & Slightly blended: N-CH$_3$NHCHO \\  
44.3536663 (46) & 15$_{4,11}$--14$_{4,10}$ & --4.9045 & 19.5 & 0.6 & 25.5 & 26.6 & 7.8 & Slightly blended: U-line  \\  
45.9549697 (61) & 18$_{1,18}$--17$_{1,17}$ & --4.6597 & 21.3 & 0.9 & 56.7 & 3.9  & 11.1 & Blended: HOCH$_2$CN \\      
45.9582741 (61) & 18$_{0,18}$--17$_{0,17}$ & --4.7687 & 21.3 & 0.9 & 33.0 & 21.2 & 6.5 & Unblended$^{*}$   \\
48.4405979 (78) & 19$_{1,19}$--18$_{1,18}$ & --4.5935 & 23.6 & 1.1 & 33.2 & 12.6 & 5.4 & Unblended$^{*}$ \\      
48.4425540 (78) & 19$_{0,19}$--18$_{0,18}$ & --4.7026 & 23.6 & 1.1 &      &               &    & Unblended$^{*}$ \\
\botrule\botrule
\end{tabular*}
\label{tab:transitions}
\footnotetext{Note: $^{(a)}$ The rotational energy levels are labeled using the conventional notation for asymmetric tops: $J_{K_{a},K_{c}}$, where $J$ denotes the angular momentum quantum number, and the $K_{a}$ and $K_{c}$ labels are projections of $J$ along the $a$ and $c$ principal axes. $^{(b)}$ We give the contribution of the (unknown) contamination to the overall area after subtracting the LTE fit of 2,5-CT from the observed spectrum, including also the contribution from all the molecules previously detected in the molecular line survey of G+0.693. $^{(c)}$ The S/N ratio is computed from the integrated signal ($\int$ $T$$\mathrm{_A^*}$d$\varv$) and noise level, $\sigma$ = rms $\times$ $\sqrt{\delta v \times \mathrm{FWHM}}$, where $\delta$$\varv$ is the velocity resolution of the spectra and the FWHM is fitted from the data. Transitions with the $^{*}$ symbol are (auto)blended with another transition of 2,5-CT. Numbers in parentheses represent the predicted uncertainty associated to the last digits.}
\end{sidewaystable}

\begin{table}
\captionsetup{labelformat=empty}
\caption{Extended Data Table 3. Total nuclear-spin rotational partition functions of 2,5- and 2,4-cyclohexadien-1-thione}             
\label{table:qr}      
\centering                          
\begin{tabular*}{0.6\textwidth}{@{\extracolsep\fill}S[table-format=3.3] S[table-format=8.4] S[table-format=8.4]}       
\midrule
{Temperature (K)} & {2,5-CT} & {2,4-CT} \\ 
\midrule 
300.0	&   8630336.0532 &  17280951.3727 \\
225.0	&   5604765.7502 &  11222535.6993 \\
150.0	&   3050479.2413 &  6107946.6382 \\
75.0	&   1078472.3004 &  2159384.3021 \\
37.5	&	 381367.1993 &  763592.2474 \\
18.75	&	 134899.6435 &  270102.2843 \\
9.375   &	  47743.8953 &  95595.2424 \\
\bottomrule
\bottomrule
\end{tabular*}
\vspace{0.7em}
\begin{flushleft}
       {Note. Statistical weights for even and odd states are 28:36 and 64:64 for 2,5-CT and 2,4-CT, respectively. Partition functions take only vibrational ground state and rotational states up to $J_{max}$=300 into account.}
\end{flushleft}
\end{table}

\end{document}